\documentclass{article}
\usepackage{spconf,amsmath,graphicx}
\usepackage{xcolor}
\usepackage{balance}

\title{Early detection of hip periprosthetic joint infections\\ through CNN on Computed Tomography images}
\makeatletter
\def\@name{ \emph{Guarnera F.$^{\star}$, Rondinella A.$^{\star}$, Giudice O.$^{\star}$, Ortis A.$^{\star}$, Battiato S.$^{\star}$},  \\ 
\emph{Rundo F.$^{\dagger}$, Fallica G.$^{\diamond}$, Traina F.$^{\S}$, Conoci S.$^{\ddagger}$}}
\makeatother

\address{$^{\star}$Department of Mathematics and Computer Science, University of Catania, Catania, Italy\\
$^{\dagger}$STMicroelectronics, ADG R\&D Power and Discretes Division, Catania, Italy  \\
$^{\diamond}$IBMTech s.r.l., via Napoli 116, Catania, Italy
\\
$^{\S}$University of Bologna, Rizzoli Orthopedic Institute of Bologna, Bologna, Italy  \\ 
$^{\ddagger}$Department ChBioFarAm, University of Messina, Messina, Italy \\\\
\emph{francesco.guarnera@unict.it, alessia.rondinella@phd.unict.it,  giudice@dmi.unict.it, }\\
\emph{ortis@dmi.unict.it, battiato@dmi.unict.it,  francesco.rundo@st.com,}\\
\emph{giorgio.fallica@ibmtech.it, francesco.traina@ior.it, sabrina.conoci@unime.it}
}

\begin{document}
%
\maketitle
\begin{abstract}

Early detection of an infection prior to prosthesis removal (e.g., hips, knees or other areas) would provide significant benefits to patients. Currently, the detection task is carried out only retrospectively with a limited number of methods relying on biometric or other medical data. The automatic detection of a periprosthetic joint infection from tomography imaging is a task never addressed before. This study introduces a novel method for early detection of the hip prosthesis infections analyzing Computed Tomography images. 
The proposed solution is based on a novel ResNeSt Convolutional Neural Network architecture trained on samples from more than 100 patients. The solution showed exceptional performance in detecting infections with an experimental high level of accuracy and F-score.
\end{abstract}
\section{Introduction}
\label{sec:intro}
\color{black}
The surgical replacement of human joints has become increasingly common in recent years as a treatment to pathologies like osteoarthritis or rheumatoid arthritis. However, the Periprosthetic Joint Infection (PJI) that unfortunately occurs around a joint implant, still represents a serious concern for patients and physicians \cite{workgroup2011new}. A PJI can lead to pain, joint dysfunction, and the need for revision surgery, which typically pose an increased potential for further complications for the patients, as well as additional costs \cite{sculco1993economic}. To address this challenge, medical imaging plays a critical role in the early and accurate detection of PJI. Traditionally, PJI detection is carried out retrospectively by means of a combination of methods relying on biometric or other medical data \cite{ting2017diagnosis} which could be time-consuming and limited in term of detection accuracy. To counter this, Machine Learning (ML) techniques could train on past data and be applied to PJI detection, with the aim of improving the accuracy of diagnosis \cite{kuo2022periprosthetic} \cite{klemt2021machine}.  Certainly, the use of Computed Tomography (CT) images provided several advantages in the diagnosis of infection \cite{cyteval2002painful}, giving that the periosteal response should be a strong prior when the infection is present \cite{kapadia2016periprosthetic}. With this hypothesis, the information obtained from CT scans could be employed as input to supervised ML approaches to develop models able to accurately predict PJIs, reducing the risk of recurrent infections and improving patient outcomes.
However, the PJI is particularly challenging to be diagnosed, even for physicians.
In this paper, a Convolutional Neural Network (CNN) solution for the classification of infected and aseptic patients with hip replacements by analyzing CT scans is presented \cite{conoci2021image}. 

The proposed solution exploits the attention mechanism of recent ResNeSt \cite{Zhang_2022_CVPR} architecture to proper extract features from CT scans and build a predictive model that can accurately differentiate between infected and non-infected (aseptic) patients. A private dataset acquired at Rizzoli Orthopedic Institute (IOR) was employed for training and final evaluation of the solution (approved by ethical committee). The dataset was manually labeled by expert radiologists. The obtained results showed the effectiveness of the proposed solution in the PJI detection task, demonstrating its potential to improve overall medical outcomes. This highlights the importance of medical imaging in the diagnosis of PJI and underscores the potential of machine learning techniques in this area.

The remainder of this paper is organized as follows: Section \ref{sec:sota} briefly presents related works; Section \ref{sec:methodology} details the proposed solution and Section \ref{sec:results} shows the experimental results with discussion. Conclusions are summarized in Section \ref{sec:future_work}.

\begin{figure}[t]
\centering
\begin{minipage}[b]{.48\linewidth}
  \centering
  \centerline{\includegraphics[width=3.8cm]{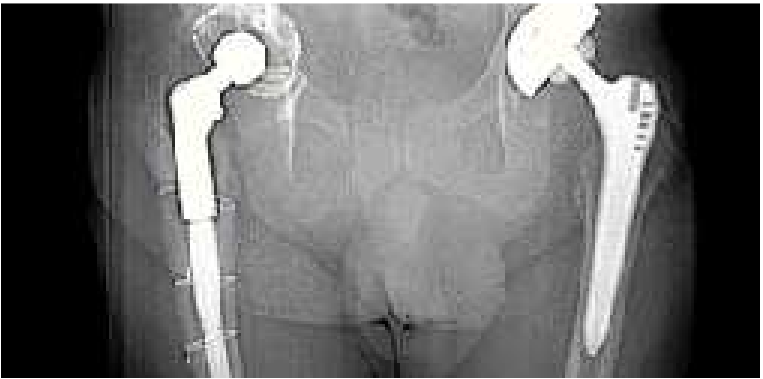}}
  \centerline{Coronal}\medskip
\end{minipage}
\begin{minipage}[b]{.48\linewidth}
  \centering
  \centerline{\includegraphics[width=3.4cm]{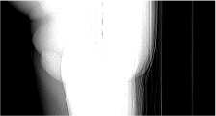}}
  \centerline{Sagittal}\medskip
\end{minipage}
\begin{minipage}[b]{.48\linewidth}
  \centering
  \centerline{\includegraphics[width=3.8cm]{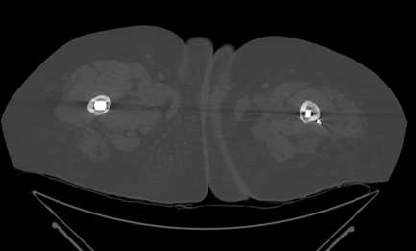}}
  \centerline{Axial Bone-Tissue}\medskip
\end{minipage}
\begin{minipage}[b]{.48\linewidth}
  \centering
  \centerline{\includegraphics[width=4.1cm]{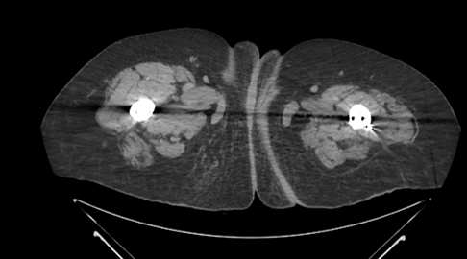}}
  \centerline{Axial Soft-Tissue}\medskip
\end{minipage}
\caption{Examples of acquisitions in CT}
\label{fig:ct_acquisitions}
\end{figure}

\section{Related works}
\label{sec:sota}
The state of the art (SOTA) methods for the detection of PJIs employ mainly statistical methods, such as regression and Fisher test. In \cite{bulow2022prediction} a risk prediction model is proposed for PJI detection within 90 days after surgery using Least Absolute Shrinkage and Selection Operator (LASSO) regression analysis \cite{tibshirani1996regression}. Other approaches propose the use of the Fisher's test, to detect the infection from CT images at the site of hip prosthesis before surgery \cite{galley2020diagnosis} \cite{isern2020value}.
While these methods demonstrated to be effective, they are limited in capturing complex relationships between imaging features thus better understanding the PJI status. For instance, many patients' problems, such as mechanical loosening, can be detected only retrospectively by studying the bone-prosthesis system and its evolution \cite{andra2008structural}. Only a limited number of studies exploited machine learning (ML) techniques. This is likely due to issue concerning datasets: availability of reliable labeling, variability in image quality and in patients samples, etc. However, ML approaches have the potential to significantly enhance the accuracy of PJI detection. 
Several studies have demonstrated the potential of deep learning approaches for PJI prediction, with promising results in terms of accuracy, sensitivity, and specificity. The authors of \cite{tao2022preliminary} proposed a ResNet architecture to classify pathological sections of patients with PJI achieving high accuracy. In \cite{klemt2021machine}, a ML approach was proposed to anticipate recurrent infections in patients who have undergone revision total knee arthroplasty due to a PJI. In some cases, the performance of these models has been shown to be superior to traditional diagnostic methods that rely solely on clinical inspection and laboratory tests. To the best of the authors' knowledge, as far as hip CT images are concerned, there are no SOTA methods able to early detect PJIs. This specific medical imaging problem is still in its early stages and there are several obstacles that need to be overcome: the most important one is the scarcity of large datasets. 

Given this, a specific dataset of hip CT scan images was collected and labeled by experts. Thus, a novel CNN solution was introduced in order to advance the state of the art in early PJI detection task, achieving the first automatic approach with promising high level of accuracy.

\section{Methodology}
\label{sec:methodology}
The CT scans are a widely used medical imaging data providing detailed cross-sectional images of internal structures of a human body. In the context of hip replacements, they play an important role in detecting a PJI. The axial plane imaging mode has the ability to highlight both bone and soft tissue (see Fig.\ref{fig:ct_acquisitions}), thus facilitating the identification of a orthopedic joint and bone infections. 

\subsection{Hip CT scans image dataset}
\label{sec:dataset}

At first, patients who underwent hip replacement surgery and subsequent CT scans for various reasons, including postoperative follow-up, suspected PJI, or routine monitoring were identified at Rizzoli Orthopedic Institute of Bologna. As infection is not easily detectable, and there is no standard protocol for medical experts to identify it. However, physicians are able to detect infection clues in proximity to the bone borders, with particular attention to the acetabular and femoral area. Finally, they are able to confirm the presence of infection by means of cultures: this is a retrospective method. In order to have a first early detection solution, a dataset consisting of $102$ CT scans of patients with hip replacements was collected ($52$ samples of an infected patient and $50$ aseptic ones). Given the before mentioned insight, only images regarding the axial bone tissue were employed. The labeling of this dataset posed a further challenge: the labels had to be assigned at patient level. In other words, each patient is labeled as either being infected or aseptic, rather than each individual CT scan image. Thus, all images from a specific patient will be labeled as infected or aseptic. Also each CT scan contains a different number of images per patient. These are big issues for ML solutions as it means that no explicit information is available about the presence or absence of a PJI on each individual images. All collected images are in Digital Imaging and Communication in Medicine (DICOM) format\footnote{https://www.dicomstandard.org/}. 

\subsection{Pre-processing pipeline}
\label{sec:preprocessing}

In order to identify what images contain the prosthesis, every pixel of the images have been converted through the Hounsfield scale, as follows: 
\begin{equation}
\label{eq:hoounsfiled}
    h_{p} =  p * s + i
\end{equation}
where $p$ is a pixel value, $s$ and $i$ are the slope and the intercept respectively contained in the DICOM metadata, and $h_{p}$ is the value of the pixel in the Hounsfield scale. The Hounsfield scale is a quantitative measure for describing radio-density which is the property of relative transparency of a material to the passage of the X-ray portion of the electromagnetic spectrum. Images with $h_{p} > 3000$ were selected as being images presenting non-human material like a metallic prosthesis \cite{morar2022analysis}. This allowed the selection of images coming from the upper part of the acetabulum to the lower part of the stem only. 

\begin{figure}[t]
\centering
\begin{minipage}[b]{\linewidth}
  \centering
  \centerline{\includegraphics[width=6.8cm]{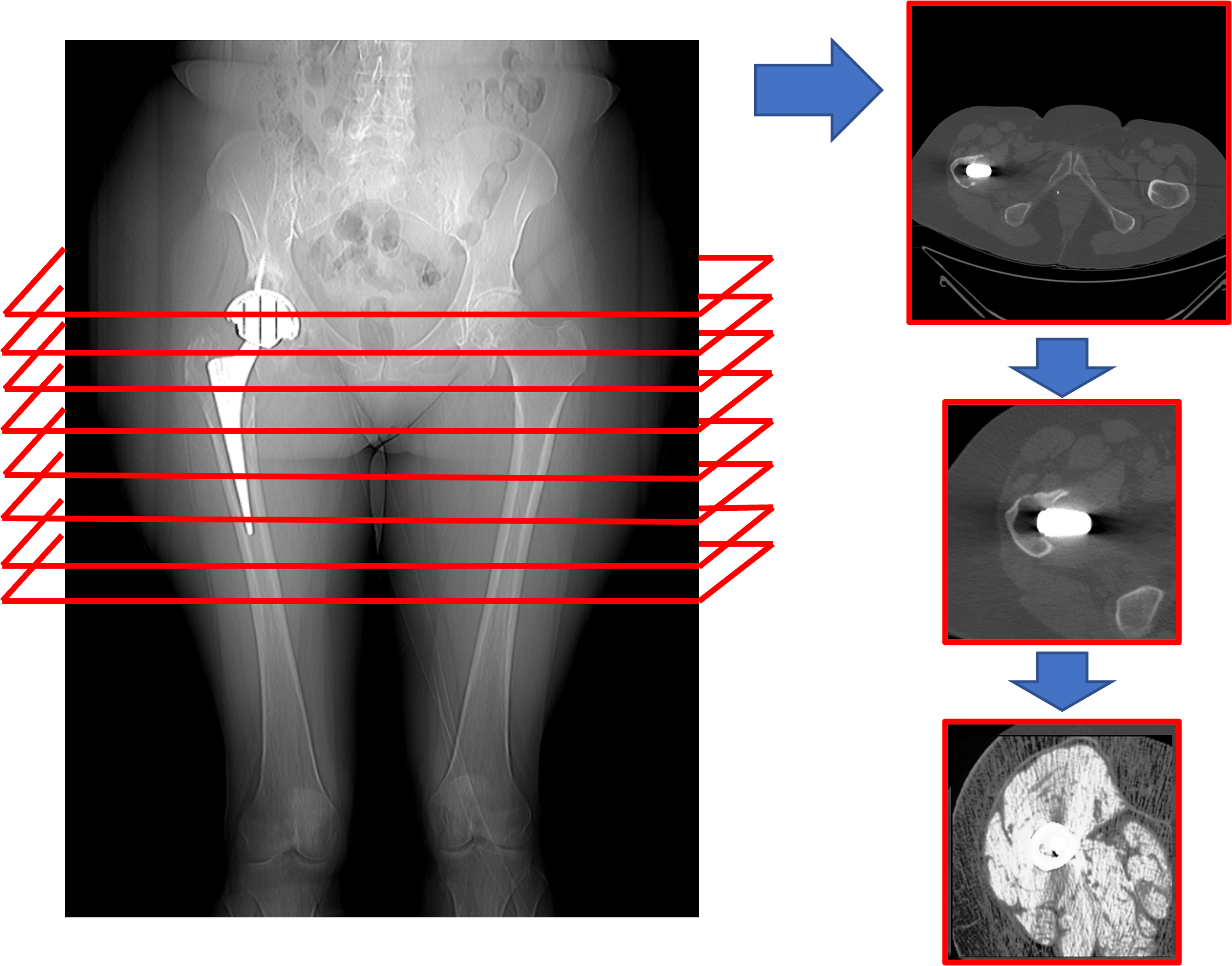}}
\end{minipage}

\caption{Pre-processing pipeline: Slice selection based on Hounsfield selection, contour-based ROI detection and final histogram equalization.}
\label{fig:preprocessing}
\end{figure}

A further preprocessing operation was carried out extracting the contours of the bone relative to the prothesis \cite{suzuki1985topological} on images having $h_{p} > 1000$. Once the contours were extracted, the corresponding centroid was computed and a sub-image of $188 \times 188$ pixels was extracted. Finally, histogram equalization was carried out on these patches to adjust the contrast and to normalize different image histograms. The overall pre-processing operations are graphically summarized in Fig. \ref{fig:preprocessing}.

\subsection{The proposed approach for PJI detection}
\label{sec:model}
In this section the proposed CNN-based solution for the early detection of hip PJI detection is described. A ResNeSt architecture \cite{Zhang_2022_CVPR} was employed to classify each data sample (CT scan images from a patient with hip replacements) between infected or aseptic classes. The ResNeSt architecture is an evolution of the ResNet neural network \cite{he2016deep}, incorporating split-attention blocks instead of residual blocks, thus emphasizing the correlation of informative feature maps. ResNeSt was selected for being the best on the specific task, after testing other similar architectures. More specifically, performances obtained by the ResNet and the ResNeXt architectures \cite{xie2017aggregated} were compared, 
However, the ResNeSt network architecture, incorporating split-attention mechanisms, demonstrated superior results in both the validation set and the test set. The other evaluated architectures, although achieving similar levels of accuracy, demonstrated poor generalization performances, as indicated by their "randomic behaviour" on heatmap analysis. The proposed approach takes input images of size $188 \times 188$ that have undergone pre-processing, as described in Section \ref{sec:preprocessing}. The images are then processed within the architecture and the loss function is computed based on the comparison between the output predictive values and truth labels of the data. The loss function employed was the cross entropy and moreover, in order to improve the stability of the model, the Jacobian regularization was applied \cite{hry2019jacobian}. Jacobian regularization permitted to reduce the impact of input perturbations caused by metallic artifacts in the prostheses. 
The use of ResNeSt with the addition of regularization techniques allowed to exploit the strengths of this state-of-the-art network architecture to extract features from CT scans and accurately classify infected and aseptic patients with hip replacements (with explainable non-randomic heatmaps).

%
%

\begin{table*}[t]
        \tiny
	\centering
	\resizebox{\linewidth}{!}{
	\renewcommand{\arraystretch}{0.98} 
		\begin{tabular}{c||c||c||c||c||c||c}
			\hline \hline
                Patient & Patient  & Number  & Accuracy / F-score &  Accuracy / F-score &  Accuracy / F-score &  Accuracy / F-score\\
                  number &  type & images & configuration $C_1$ & configuration $C_2$ & configuration $C_3$ & configuration $C_4$\\
			\hline 
                1 & aseptic & 67 & 0.88/0.94 & 0.88/0.93 & 0.91/0.95 & 0.89/0.94\\
                \hline 
                2 & aseptic & 39 & 0.66/0.8 & 0.41/0.58 & 0.92/0.96 & 0.56/0.72\\
                \hline 
                3 & aseptic & 69 & 0.85/0.85 & 0.76/0.87 & 0.76/0.87 & 0.67/0.8\\
                \hline 
                4 & aseptic & 68 & 0.52/0.69 & 0.85/0.92 & 0.70/0.83 & 0.57/0.73\\
                \hline 
                5 & aseptic & 77 & 0.79/0.88 & 0.79/0.88 & 0.96/0.98 & 0.82/0.9\\
                \hline 
                6 & aseptic & 53 & 0.96/0.98 & 1/1 & 1/1 & 0.92/0.96\\
                \hline 
                7 & infected & 80 & 0.46/0.63 & 0.51/0.68 & 0.39/0.55 & 0.65/0.79\\
                \hline 
                8 & infected & 115 & 1/1 & 1/1 & 0.98/0.99 & 1/1\\
                \hline 
                9 & infected & 191 & 0.98/0.99 & 0.91/0.96 & 0.82/0.90 & 0.92/0.96\\
                \hline 
                10 & infected & 71 & 0.18/0.31 & 0.13/0.20 & 0.03/0.05 & 0.13/0.22\\
                \hline 
                11 & infected & 77 & 0.57/0.72 & 0.89/0.94 & 0.52/0.68 & 0.91/0.95\\
                \hline 
                12 & infected & 117 & 0.95/0.98 & 0.96/0.98 & 0.99/0.99 & 0.91/0.95\\

                \hline \hline

			\hline
		\end{tabular}
	}
	\caption{Accuracy and F-score obtained on $C_1$,$C_2$,$C_3$,$C_4$ configurations employing the same test set $D_{x}$.}
	\label{tab:result_tables}
\end{table*}

\section{Experiments and results}
\label{sec:results}
As outlined in Section \ref{sec:dataset}, the dataset collected and employed for experiments is composed of $50$ aseptic and $52$ infected patients. To evaluate the generalizing properties of the proposed solution, a balanced test set ($D_{x}$) of $12$ patients ($6$ aseptic and $6$ infected) was extracted from the full dataset, and four other balanced different validation sets ($D_{v1}$, $D_{v2}$, $D_{v3}$, $D_{v4}$) each consisting of different $12$ patients. The remaining $42$ patients were considered as $D_{t}$. The following configurations were considered for the training phase:

\begin{itemize}
    \item $C_1$: train \{$D_{t} \cup D_{v2} \cup D_{v3} \cup D_{v4}$\}  and valid  $D_{v1}$;
    \item $C_2$: train  \{$D_{t} \cup D_{v1} \cup D_{v3} \cup D_{v4}$\} and valid  $D_{v2}$;
    \item $C_3$: train \{$D_{t} \cup D_{v1} \cup D_{v2} \cup D_{v4}$\} and valid  $D_{v3}$;
    \item $C_4$: train \{$D_{t} \cup D_{v1} \cup D_{v2} \cup D_{v3}$\} and valid  $D_{v4}$.
\end{itemize}

\begin{figure}[t]
\centering
\begin{minipage}[b]{\linewidth}
  \centering
  \centerline{\includegraphics[width=8.5cm]{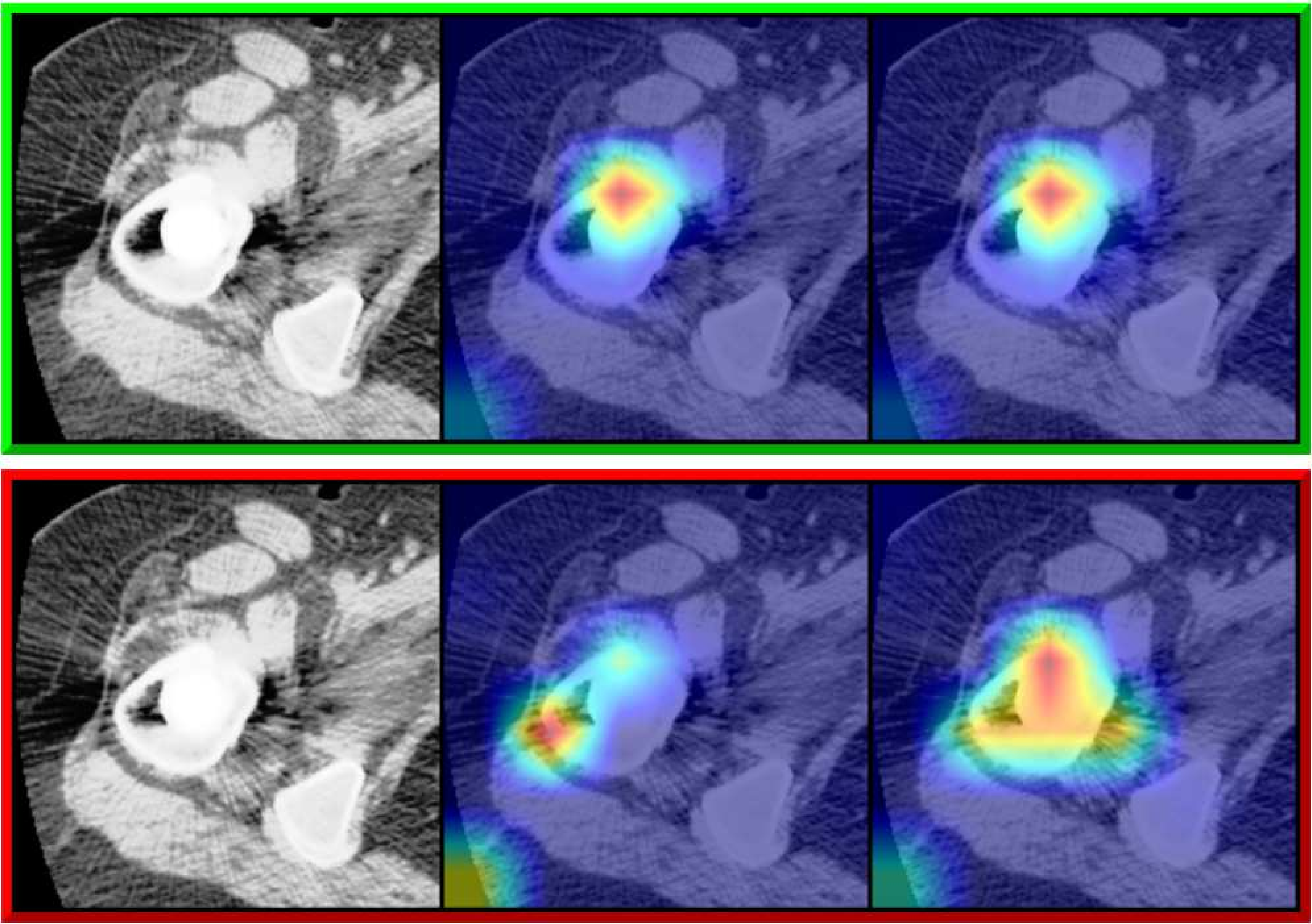}}
\end{minipage}
\caption{Input images and corresponding heatmaps of trained model attention produced by means of GradCam and GradCam++ libraries. Green border indicates a correct prediction while the red border shows a wrong one.}
\label{fig:gradcam}
\end{figure}

Training was performed for $100$ epochs, employing ADAM as optimizer with a starting learning rate of $1e-4$, a weight decay equal to $1e-4$ and a batch size fixed at $4$. The proposed approach was implemented in Python language (version 3.9.7) using the Pytorch package. All experiments were done on a NVIDIA Quadro RTX 6000 GPU. The overall training procedure took 50 hours.

Table \ref{tab:result_tables} shows the results obtained applying the best model obtained for each configuration and tested on $D_{x}$. As explained in Section \ref{sec:model} the proposed technique makes a classification (positive or infected vs. negative or aseptic) for each image of a patient; this implies that the model is independent to the number of images per patient, as each patient has a different number of images based on their length of the prosthesis (third column of Table \ref{tab:result_tables}). It has to be noted that, in order to accurately classify a patient as infected or aseptic through the images predicted by the model, a threshold strategy must be employed. The choice of the threshold could be done considering the right trade-off between false positive and false negative.

Accuracy (Eq. \ref{eq:accuracy}) and F-score (Eq. \ref{eq:f1}) were employed as performance metrics as shown in Table \ref{tab:result_tables}.
Accuracy is defined as follows:

\begin{equation}
\label{eq:accuracy}
    ACC_{y} =  Y_{rp} / PNI 
\end{equation}

where $y$ is the class, $Y_{rp}$ represents the right prediction for that class and $PNI$ is the number of images per patient.
F-score is defined as follows:

\begin{equation}
\label{eq:f1}
    F =  2* (P*R)/(P+R) 
\end{equation}

where $P$ and $R$ are precision and recall respectively. 
Results shown in Table \ref{tab:result_tables} show how the model generalizes with respect to the training set: accuracy and F-score are similar for each one of the cross-validation configurations. 

Experts treating the task of PJI detection advise on focusing on regions on images adjacent to the bones, which is often the discriminating factor. To furtherly verify this expert knowledge, a Gradient-weighted Class Activation Mapping (Grad-Cam) \cite{selvaraju2017grad} analysis was carried out on each test image with the trained model. This analysis was performed on both Grad-Cam available versions, and the results showed that if the model highlights discriminative features near the bones, the prediction is accurate (the green-bordered example in Fig. \ref{fig:gradcam}). On the other hand, if the model concentrated on multiple features of the image, the prediction was incorrect. 

\section{Conclusion and future works}
\label{sec:future_work}
In this paper an early detection technique for hip periprosthetic joint infections on computed tomography images was presented. This was a first in the state of the art to the best of author's knowledge. To this aim, hip CT images were collected and labeled by experts.
A dedicated pre-processing pipeline was developed and a ResNeSt CNN solution was trained. The proposed pipeline demonstrated to achieve strong performances (in terms of accuracy and F-score) in detecting infected and aseptic images of a specific patient. 

Future research will explore alternative forms of input, such as compressed images, and the implementation of additional automated techniques to further enhance the prediction of prosthesis-related joint infections. Additionally, we will investigate the outliers in the collected dataset to better understand their impact on the overall results.

\subsection*{Acknowledgements}
\label{sec:acknowledgements}
Alessia Rondinella is a PhD candidate enrolled in the National PhD in Artificial Intelligence, XXXVII cycle, course on Health and life sciences, organized by Università Campus Bio-Medico di Roma.

Experiments were carried out thanks to the hardware and software granted and managed by iCTLab S.r.l. - Spinoff of University of Catania (https://www.ictlab.srl).

\vfill\pagebreak

\balance
\bibliographystyle{IEEEbib}
\bibliography{refs}

\end{document}